\documentclass[twocolumn,showpacs,preprintnumbers,amsmath,amssymb,superscriptaddress]{revtex4}
\usepackage{epsf}
\usepackage{graphicx}
\usepackage{sidecap}

\usepackage{color}

\def \beq {\begin{equation}}
\def \eeq {\end{equation}}
\begin{document}

\title{{Surface electronic structure of a topological Kondo insulator candidate SmB$_6$: insights from high-resolution ARPES}}
\author{M.~Neupane}
\affiliation {Joseph Henry Laboratory and Department of Physics,
Princeton University, Princeton, New Jersey 08544, USA}

\author{N.~Alidoust}\affiliation {Joseph Henry Laboratory and Department of Physics, Princeton University, Princeton, New Jersey 08544, USA}

\author{S.-Y.~Xu}
\affiliation {Joseph Henry Laboratory and Department of Physics, Princeton University, Princeton, New Jersey 08544, USA}

\author{T.~Kondo}
\affiliation {ISSP, University of Tokyo, Kashiwa, Chiba 277-8581, Japan}

\author{D.-J. Kim}
\affiliation {Department of Physics and Astronomy, University of California at Irvine, Irvine, CA 92697, USA}

\author{Chang~Liu}
\affiliation {Joseph Henry Laboratory and Department of Physics,
Princeton University, Princeton, New Jersey 08544, USA}

\author{I.~Belopolski}\affiliation {Joseph Henry Laboratory and Department of Physics, Princeton University, Princeton, New Jersey 08544, USA}

\author{T.-R. Chang}
\affiliation{Department of Physics, National Tsing Hua University, Hsinchu 30013, Taiwan}

\author{H.-T. Jeng}
\affiliation{Department of Physics, National Tsing Hua University, Hsinchu 30013, Taiwan}
\affiliation{Institute of Physics, Academia Sinica, Taipei 11529, Taiwan}

\author{T.~Durakiewicz}
\affiliation {Condensed Matter and Magnet Science Group, Los Alamos National Laboratory, Los Alamos, NM 87545, USA}

\author{L. Balicas}
\affiliation {National High Magnetic Field Laboratory, Florida State University, Tallahassee, Florida 32310, USA}
\author{H.~Lin}
\affiliation {Department of Physics, Northeastern University,
Boston, Massachusetts 02115, USA}
\author{A.~Bansil}
\affiliation {Department of Physics, Northeastern University,
Boston, Massachusetts 02115, USA}
\author{S.~Shin}
\affiliation {ISSP, University of Tokyo, Kashiwa, Chiba 277-8581, Japan}
\author{Z. Fisk}
\affiliation {Department of Physics and Astronomy, University of California at Irvine, Irvine, CA 92697, USA}
\author{M.~Z.~Hasan}
\affiliation {Joseph Henry Laboratory and Department of Physics,
Princeton University, Princeton, New Jersey 08544, USA}

\date{18 June, 2013}
\pacs{}
\begin{abstract}

\textbf{The Kondo insulator SmB$_6$ has long been known to exhibit low temperature (T $<$ 10K) transport anomaly and has recently attracted attention as a new topological insulator candidate. By combining low-temperature and high energy-momentum resolution of the laser-based ARPES technique, for the first time, we probe the surface electronic structure of the anomalous conductivity regime. We observe that the bulk bands exhibit a Kondo gap of 14 meV and identify in-gap low-lying states within a 4 meV window of the Fermi level on the surface of this material. The low-lying states are found to form electron-like Fermi surface pockets that enclose the \textbf{X} and the $\mathbf{\Gamma}$ points of the surface Brillouin zone. These states disappear as temperature is raised above 15K in correspondence with the complete disappearance of the 2D conductivity channels in SmB$_6$. While the topological nature of the in-gap metallic states cannot be ascertained without spin (spin-texture) measurements our bulk and surface measurements carried out in the transport-anomaly-temperature regime (T $<$ 10K) are consistent with the first-principle predicted Fermi surface behavior of a topological Kondo insulator phase in this material.}

\end{abstract}
\date{\today}
\maketitle

Materials with strong electron correlations often exhibit exotic ground states such as the heavy fermion state, Mott insulator, Kondo insulator and related superconductors. Kondo insulators are mostly realized in the rare-earth materials featuring \textit{f}-electron degrees of freedom, which behave like a correlated metal at high temperature, with a bulk band gap opening at low temperature \cite{Fisk, Coleman, Riseborough}. The energy gap is attributed to hybridization of the nearly localized, flat 4\textit{f} bands located near the Fermi level with the dispersive conduction band. With the advent of 3D topological insulators \cite{Hasan, Hasan2, Hsieh, SCZhang, Xia} the compound SmB$_6$, sometimes categorized as heavy-fermion semiconductor \cite{Fisk, Coleman, Riseborough}, attracted much attention due to the theoretical proposal that it may possibly host the topologically protected states within its Kondo bandgap \cite{Dzero,Takimoto,Dai}. This compound is metallic at room temperature, but the resistivity goes up at low temperatures and saturates to a finite value below 6K \cite{Menth, Allen, Cooley}. The anomalous residual conductivity present in SmB$_6$ at low temperature is believed to be associated with the states that lie within the Kondo bandgap, indirectly evidenced in various experiments \cite{Kimura, Nanba, Nyhus, Alekseev, point_cont, Miyazaki, Denlinger}.

 From a topological insulator perspective, the low temperature resistivity anomaly was attributed to topological surface states \cite{Dzero,Takimoto,Dai}. Several surface-related transport phenomena have recently been revealed in SmB$_6$ \cite{Fisk_discovery, Hall, tunelling} providing further insights into the materials 2D conduction below 10K. However, a direct experimental evidence of temperature dependent evolution of surface states and their momentum resolved structure and anisotropy in this transport regime has not been achieved to this date. Previous ARPES studies have been limited to the temperature regime of 15-20K or higher \cite{ Miyazaki, Denlinger} with limited resolution or perhaps sample quality. We perform high-resolution and low temperature ARPES measurements to illuminate on the nature of residual conductivity anomaly below 10K from a band structure evolution perspective, and to test the proposal of the presence of conducting surface states within the Kondo bandgap of SmB$_6$. Additionally, systematic temperature dependence is carried out throughout the Kondo transport regime to determine the size of the Kondo gap and the \textit{d-f} hybridization process and their relation to the transport anomaly.
Experimentally, at low temperatures we identify low-lying states near the Fermi level which vanish above 15K. We observe clear signatures of bulk band hybridization at low temperatures featuring a Kondo gap of about 14 meV, as well as the in-gap states within 4meV from the Fermi level. These states, as well as the transport surface conductivity or the 2D anomaly disappear above 15K. The detailed work presented here lays the foundation to understanding the residual conduction behavior and transport anomaly of SmB$_6$ universally observed at low temperatures (T $<$ 10K).


Single crystalline samples of SmB$_6$ used in our measurements were grown by Al-flux method in Fisk Lab at University of California (Irvine) which is detailed elsewhere \cite{Fisk_discovery, Hall}.
Synchrotron-based ARPES measurements of the low-energy electronic structure were performed at the Synchrotron Radiation Center (SRC), Wisconsin, equipped with high efficiency R4000 electron analyzer using PGM beamline.
Separate, high-resolution and low temperature ARPES measurements were performed using a Scienta R4000 hemispherical analyzer with an ultraviolet laser ($h \nu = 6.994$eV) at the Institute for Solid State Physics (ISSP) at Tokyo.
The energy resolution was set to be 4 meV and 20 meV for the measurements with the laser source or the synchrotron beamline, respectively, and angular resolution was set to be better than 0.1$^{o}$ for all measurements.
The first-principles calculations were based on the generalized gradient approximation (GGA) \cite{GGA} using the projector augmented-wave method \cite{PAW} as implemented in the VASP package \cite{VASP}. The experimental crystal structure was used \cite{expt} for calculations. The spin-orbit coupling was included self-consistently in the electronic structure calculations with 12$\times$12$\times$12 Monkhorst-Pack $k-$mesh.


We start out presenting crystal symmetry and the Brillouin zone structure of SmB$_6$ materials, which crystallize in the CsCl-type structure with Sm ions and B$_6$ octahedron being located at the corner and body center of the cubic lattice, respectively (see Fig. 1a). The bulk and projected Brillioun Zone (BZ) of (001) surface for SmB$_6$ are shown in Fig. 1(b). The resistivity of our sample shows a sharp increase at T$<$ 30K and start to saturate at T$<$ 7K (Fig. 1c) in agreement with previous reports \cite{Menth, Cooley}.
The resistivity versus temperature is plotted in Fig. 1c for the sample characterization.
In Figs. 1d and 1e, we present ARPES intensity plots measured along  ${\textrm{M}}-{\textrm{X}}-{\textrm{M}}$ and ${\textrm{X}}-\Gamma-{\textrm{X}}$ directions over the wide binding energy scale measured with the synchrotron-based ARPES system with photon energy of 26 eV at temperature of 12 K. The non-dispersive Sm 4$f$ states near the Fermi level are observed which are also marked in the integrated energy distribution curves (EDCs) plotted next to the intensity plots. The position of the flat bands are estimated to be at binding energy (BE) of 15 meV, 150 meV and 1 eV. These states are assigned as $^6$H$_{5/2}$, $^6$H$_{7/2}$ and $^6$F, which are multiples of the Sm$^{2+}$ ($f^6$ to $f^5$) final states, in agreement with previous report \cite{Miyazaki}. More importantly, we observe a dispersive parabolic band centered at $\bar{\textrm{X}}$ point, which come from Sm 5$d$ band. The hybridization between the Sm 5$d$ band and Sm 4$f$ flat band is clearly visible especially at binding energy of 150 meV in Figs. 1d and 1e.

In order to search for the temperature dependent in-gap states, we employ the high-resolution low temperature laser-based ARPES system. Figs. 2a and 2b show a schematic scenario  of temperature dependent hybridization and its integrated density of states. The density of states measured below hybridization temperature shows a small dispersive peak crossing the Fermi level which vanishes above the hybridization temperature. Guided by recent first-principle calculations predicting in-gap surface states around $\Gamma$ and $\textrm{X}$ points \cite{Dai,Takimoto}, we focus our detailed temperature dependent measurement around these high-symmetry points. In Fig. 2c, we show the integrated EDCs measured at various temperatures at the band centered around the $\textrm{X}$ point. As the temperature rises, the gap between hybridized states decreases and finally the two states merge into a single state whose traced cannot be observed at 30K.  At low temperature, the gap value between hybridized states is estimated about 14 meV. We observe that the $f$-$d$ hybridization evolves with temperature, thus providing the temperature-dependent hybridization evidenced in electronic structure. 
It is also important to note that at 7eV photoionization cross section for the $f$ orbitals is negligible, so the measured states are significantly corresponding to the partial $d$-orbital character of the hybridized band.




Surprisingly, a similar temperature-dependent behavior is observed around $\Gamma$ point where no direct hybridization between $d$ and $f$ bands is seen or expected (Fig. 2d). These states are clearly the in-gap states, states inside the Kondo bandgap, and they cannot be traced in the bulk band calculations. However, they are allowed to appear due to the bulk band inversion at $\Gamma$ point (driven by strong spin-orbit interaction strength of \textit{df}-hybrid bands) which is the hallmark of topological non-triviality \cite{Dzero, Dai}. Fig. 3a shows a Fermi surface map of SmB$_6$ measured at temperature of 7K with laser energy of 7eV (see supplementary information for detailed data). A large oval-shaped and a small nearly circular-shaped Fermi surfaces (FS) are observed at $\textrm{X}$  and $\Gamma$ point, respectively, which is in a qualitative agreement with the FS calculated from first-principles \cite{Dai}. Here we note that the bulk electronic dispersion is not expected to give rise to such a FS. We conclude that the observed FS is most probably coming from the hybridized surface states due to their correspondence with surface state calculations and their absence in bulk band calculations.
The key signatures of topological Kondo insulators, such as the existence of a metallic surface state characterized by Dirac cones with helical spin-textures, are the same as those of other 3D topological insulators \cite{Hasan, Hasan2, HsiehSci, HsiehNat}. While the laser light source used here provides ultra-high energy resolution, it is limited by single photon energy and (lack of) spin resolution while also maintaining the same low temperature and ultra-high resolution conditions simultaneously. Therefore, at present such studies are not feasible. Present study thus focuses on testing the proposal of in-gap states and their correlation with low-temperature transport regime below 15K and comparison of the k-space data with first-principle calculations of topological surface states.


We plot ARPES intensity spectra for $\Gamma$ and $\textrm{X}$ band  in Fig. 3b and 3c. In both cases, the gap of about 13 or 14 meV is observed, which is clearly observed in EDCs and 2$^{nd}$ derivative plots. The comparison of the integrated EDCs for $\Gamma$ and $\textrm{X}$ bands is shown in Fig. 3d, which further confirms the appearance of small density of states near the Fermi level at both points. The observation of gap at $\textrm{X}$  point is attributed to the  hybridization between 4$f$ and 5$d$ bands, vanishing clearly by reaching 30K (Fig. 2c). The detailed interpretation of the gap observed at $\Gamma$ point is not straightforward, despite our best efforts of utilizing 4 meV resolution working at 6K temperature, since without the observation of clear connection of these states with bulk valence and conduction band, and establishing the time-reversal invariant behavior of it, these states cannot yet be called topological based on experimental data alone. Fig. 4 shows the calculated bulk band structure of SmB$_6$, which qualitatively agrees with the experimental observations. The GGA calculations give an insulating ground state with a small energy gap of 15 meV (Fig.4b). From the orbital-decomposed band structure (Fig.4c), we find that the flat Sm 4$f$ bands are located around E$_F$ from -0.5 eV. An itinerant Sm 5$d$ band with larger dispersion hybridizes with 4$f$ bands, forming an anti-crossing band shape. The agreement between low-temperature ARPES data including the k-space maps of Fermi surfaces and Kondogap value (15 meV) with our GGA calculations and previous first-principle calculations provide important insights into the surface electronic structure of  SmB$_6$.


In conclusion, the topological Kondo insulator candidate SmB$_6$ has long been known to exhibit anomalous low temperature (10K) conductivity anomaly. By combining low-temperature capability and high energy resolution of laser-based ARPES technique, for the first time we accessed this transport regime with spectroscopy. The $4f-5d$ hybridized states are observed with a Kondo bandgap of about 14 meV. Importantly, the hybridized states around at $\textrm{X}$ point vanish at high temperatures. Furthermore, the similarly temperature-dependent low-lying states are observed at $\Gamma$ point. The simultaneous presence of these states at low temperature both around $\textrm{X}$ and $\Gamma$ points with a 4 meV window of the Fermi level is responsible for the residual conducting behavior of SmB$_6$. The in-gap states are found to enclose the \textbf{X} and the $\mathbf{\Gamma}$ points of the surface Brillouin zone k-space. While the topological nature of the in-gap states cannot be ascertained without Spin-ARPES measurements our low-temperature ARPES data taken in the transport anomaly regime in SmB$_6$ help understand the long-standing puzzle of residual conduction in SmB$_6$.



\bigskip
\bigskip

\*Correspondence and requests for materials should be addressed to M.Z.H. (Email: mzhasan@princeton.edu).

\newpage

\begin{figure*}
\centering
\includegraphics[width=16.5cm]{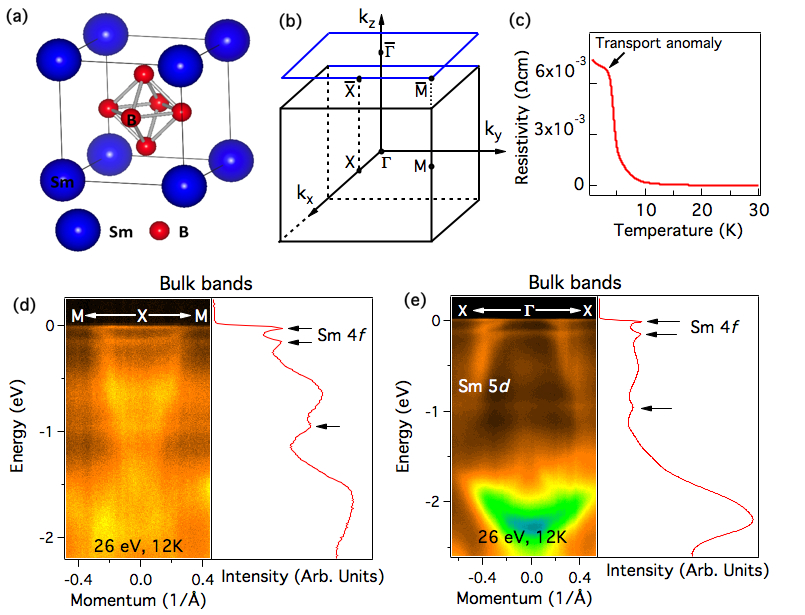}
\caption{\textbf{Crystal structure and sample characterization of SmB$_6$}. (\textbf{a}) Crystal structure of SmB$_6$. Sm ions and B$_6$ octahedron are located at the corner and centre of the cubic lattice structure.
(\textbf{b}) The bulk and surface Brillouin zones of SmB$_6$. High-symmetry points are marked.
 (\textbf{c}) Resistivity versus temperature for SmB$_6$.
  (\textbf{d}) and (\textbf{e}) Synchrotron-based ARPES experimental results: Dispersion mapping along  ${\textrm{M}}-{\textrm{X}}-{\textrm{M}}$ cut in (d) and ${\textrm{X}}-\Gamma-{\textrm{X}}$ cut in (e). Dispersive Sm 5$d$ band and non-dispersive flat Sm 4$f$ bands are observed. The integrated energy distribution curves (EDC) are also plotted to highlight the flat bands of Sm 4$f$.}
\end{figure*}


\begin{figure*}
\centering
\includegraphics[width=16.5cm]{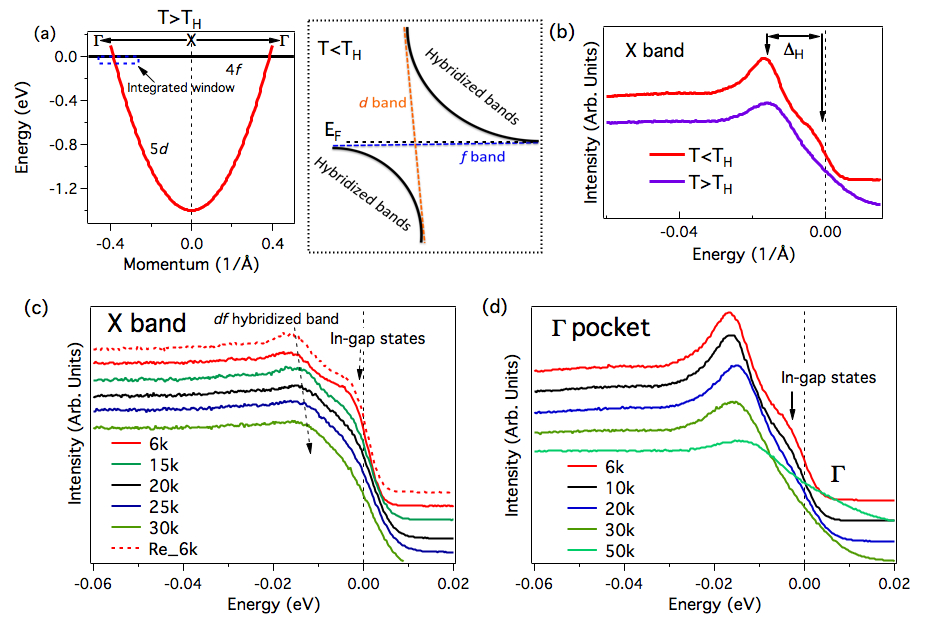}
\caption{\textbf{Temperature dependent hybridized states.}
(\textbf{a}) Schematic view of Sm 4$f$ (flat) and Sm 5$d$ (dispersive parabolic) bands above the hybridization temperature. The small black dashed rectangle represents the region of hybridization shown at  T$<T_H$. The big black dashed rectangle  shows the bands after hybridization at T$<T_H$. The solid black rectangle shows thae approximate measurement window and black dash lines represent the Fermi level (E$_F$). (b) The integrated density of states measured above and below the hybridization temperature clearly showing the hybridized states and hybridization gap.
(\textbf{c}) Temperature dependent measurements: the energy distribution curve measured at ${\textrm{X}}$ band indicated in Fig (a). Measured temperatures are also noted on the plot.
(\textbf{d}) Same as in (c) for $\Gamma$ point. The thermally recycled EDC is also plotted which confirms that these hybridization states are robust at low temperature and reproducible even after thermal cycling of the sample.}
\end{figure*}


\begin{figure*}
\centering
\includegraphics[width=16.5cm]{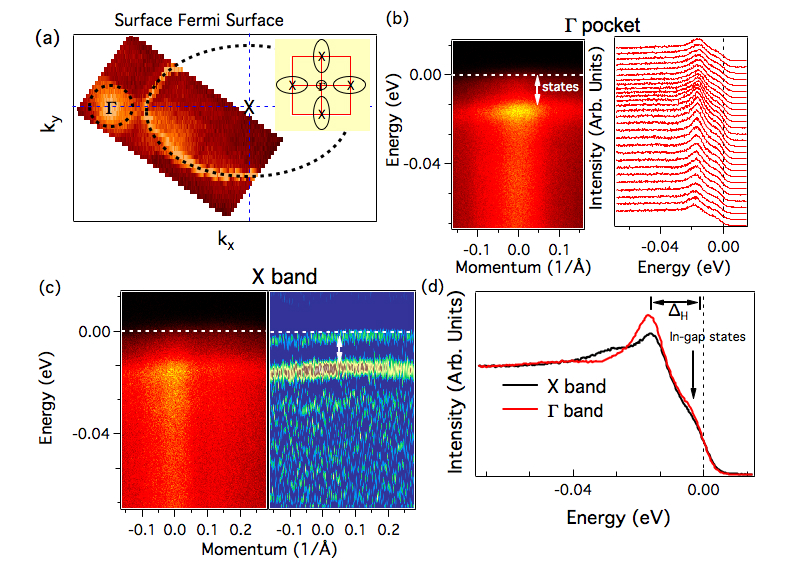}
\caption{\textbf{Fermi surface and dispersion maps of SmB$_6$.}
(\textbf{a}) Fermi surface plot of SmB$_6$ measured by 7 eV LASER source at temperature of 7 K. A small $\Gamma$ pocket and a large ${\textrm{X}}$ pocket are observed. A large elliptical and a small circular shaped black dash lines around ${\textrm{X}}$ and $\Gamma$ points are guides to the eye. Inset shows a schematic plot of Fermi surface in the first Brillouin zone. (\textbf{b}) Electronic dispersion map (left) and its energy distribution curves (EDCs) for $\Gamma$  pocket. (\textbf{c}) same as (\textbf{b}) for ${\textrm{X}}$ band. (\textbf{d}) Comparison of integrated EDC for $\Gamma$  and ${\textrm{X}}$ band. A gap value of about 15 meV is observed in both cases.}
\end{figure*}


\begin{figure*}
\centering
\includegraphics[width=16.5cm]{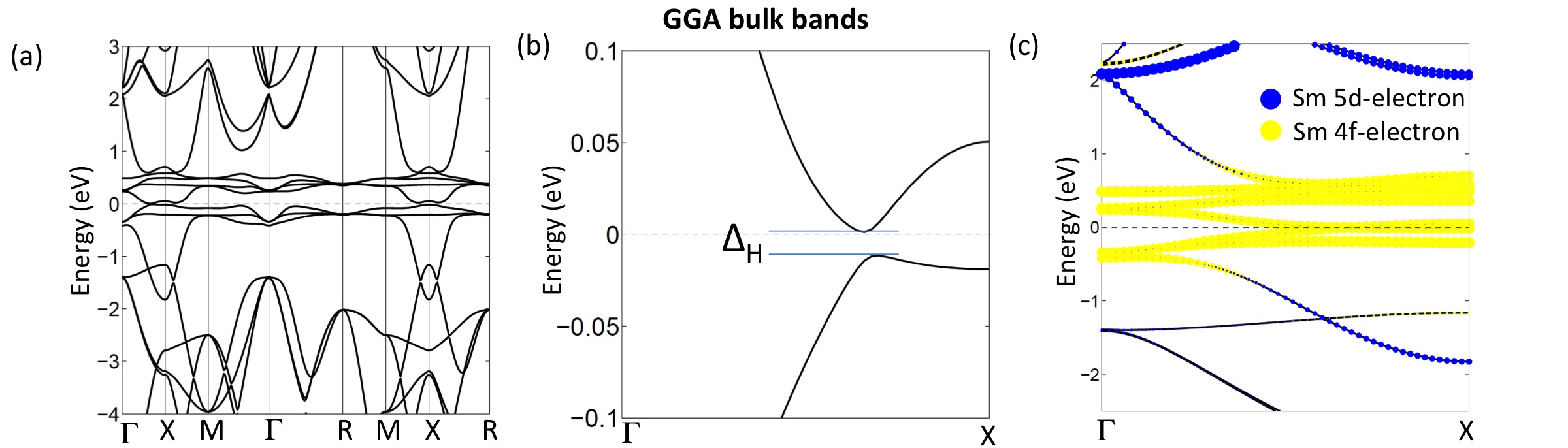}
\caption{\textbf{Calculated bulk band structure of SmB$_6$.}
(\textbf{a}) The calculated band structure of SmB$_6$ by GGA. (\textbf{b}) Zoom in of (a) near Fermi level along $\Gamma-{\textrm{X}}$. The band gap of about 15 meV was obtained in the calculation.
(\textbf{c}) Orbital decomposed band structure near Fermi level along $\Gamma-{\textrm{X}}$. The size of blue and yellow sphere is proportional to the weight of Sm 5$d$ and 4$f$ orbital, respectively.}
\end{figure*}

\end{document}